# Advanced material system for the design of an intermediate band solar cell: type-II CdTe quantum dots in a ZnCdSe matrix

V. Deligiannakis[1,3], G. Ranepura[1], I. L. Kuskovsky[2,3], M. C. Tamargo[1,3]

[1]City College of CUNY, 160 Convent Ave., New York, New York 10031, USA
[2]Queens College of CUNY, 65-30 Kissena Blvd, Queens, New York 11367, USA
[3]The Graduate Center of CUNY, 365 Fifth Avenue, New York, New York 10016, USA
mtamargo@ccny.cuny.edu

**Abstract:** We explore CdTe fractional monolayer quantum dots (QDs) in a ZnCdSe host matrix for potential application in an intermediate band solar cell device. Careful consideration has been taken during the initiation of the growth process of QDs by migration enhanced epitaxy, in order to avoid the formation of undesirable interfacial layers that can form due to the lack of common anion between the two materials. A superlattice structure of 100 periods of alternating QD and spacer layers is analyzed by high resolution X-ray diffraction (XRD) and photoluminescent (PL) spectroscopy. Simple arguments are used following continuum elastic theory to deduce the size of the dots and the strain within the superlattice from XRD data. This is further verified using PL and used in the energy calculations that yield the values of the intermediate band energy. The results suggest that the optimized materials are highly suitable for these high efficiency solar cells.

**Introduction**:

The goal of higher efficiency in solar cells has been driving solar cell research since their inception, and its significance has never been greater than it is today. One possible solution to overcoming the single junction solar cell efficiency imposed by the Schockly-Quisser limit[1] of around 31% is by introducing an intermediate band (IB) within a host material, as in the so called intermediate band solar cell (IBSC)[2]. Using either epitaxial quantum dots (QDs) or impurities one can introduce such a band[3,4]. A two step-photon process occurs from the valence band (VB) to the IB and from IB to the conduction band (CB), ultimately increasing light absorption of the solar spectrum without compromising the open circuit voltage ($V_{OC}$) of the device[5]. The proof of concept for an IBSC has been realized, and it has been shown that the short circuit current can be increased by the introduction of QDs[6]. However, with most materials used so far there is a significant drop in the $V_{OC}$, ultimately limiting the overall efficiency of the device with regards to a reference solar cell[7].

Type-II Zn(Cd)Te /ZnCdSe submonolayer QDs have been explored by our group for their promising properties as IBSCs[8,9,10]. The ZnCdSe host material when lattice matched to InP has a bandgap of ~2.1 eV, in which the Zn(Cd)Te QDs can form an intermediate band with an energy 0.3 – 0.7 eV above the VB edge. The similarity of these parameters with those required for an ideal IBSC[11] makes this material system an outstanding candidate. However, it was recently shown[12] that at the interface between the host material of ZnCdSe and the QDs an unintentional highly strained ZnSe interfacial layer is formed. If this is not accounted for, the strain accumulation in the thick stacked QD superlattice can be significant enough to lead to the formation of defects affecting the device performance. As is the case for QDs grown by the Stranski-Krastanov (SK) method, which require the formation of a strained wetting layer, as the QD layers are repeated multiple times the strain in the structure increases, decreasing the overall quality of the material[13,14,15]. The presence of an interfacial layer can also affect the bandstructure of the device. Our group recently showed that using a new shutter growth sequence we can significantly suppress the formation of the interfacial layer[16]. Now that we are able to control the interface formation, we set out to explore new QD compositions and their potentially advantageous properties. Here we pursue a new material system, based on sub-monolayer CdTe QDs embedded in the ZnCdSe host material. Besides providing a platform in which the ZnSe interfacial layer is more fully suppressed, this system has several advantages over the ZnCdTe QD system previously studied. Two main advantages are 1) the binary composition of the QD which makes it more easily controlled and more uniform, and 2) the large compressive strain in the QDs which produces a large valence band offset with respect to the matrix material (ZnCdSe) that can be exploited for better devices via band structure engineering. This strain can be easily offset by strain compensation in the spacer regions.

**Growth:**

A schematic of the structure investigated is shown in Fig. 1(a). A CdTe/ZnCdSe QD superlattice (SL) is grown by a combination of conventional molecular beam epitaxy (MBE) and migration enhanced epitaxy (MEE). The initial buffer layers and spacer material (ZnCdSe) were grown by MBE, whereas to achieve the formation of sub- monolayer CdTe QDs the MEE process was implemented.

The sample was grown on (001) oriented InP substrates with a 100 nm lattice matched InGaAs buffer layer in a dual chamber MBE growth system. The substrate and buffer layer were transferred to the II-VI chamber via ultra-high vacuum transfer modules. The II-VI layer structure included a 100 nm ZnCdSe buffer layer, followed by a superlattice consisting of alternating 13 to 14 monolayers of the ZnCdSe spacer layer and the submonolayer CdTe QDs. The formation of CdTe QDs by MEE was achieved by using a special shutter sequence of alternating Cd and Te fluxes with short wait times between them.

The shutter sequence recently developed by our group entailed a growth interruption with exposure of only a Cd flux for 5 s after the growth of the ZnCdSe spacer, followed by a short wait time of 1 s without any shutters open, instead of the 5 s wait times previously performed.[12] We have found that terminating the surface with only Cd rather than with both group II elements of Zn and Cd is an important and critical step for the avoidance of the IF layer formation[15]. The surface termination of a metal rich surface is observed by a change from a VI-rich (2×1) terminated RHEED pattern, observed during the ZnCdSe spacer layer growth, to a II-rich c(2×2) pattern after the Cd-only exposure. To grow our CdTe QDs we employed the same initial Cd-only exposure (5 s) and 1 s wait time without any impingent fluxes. After this the Cd shutter was opened for 5s, and then closed and followed by a 1s wait time. This sequence was repeated three times (three cycles).

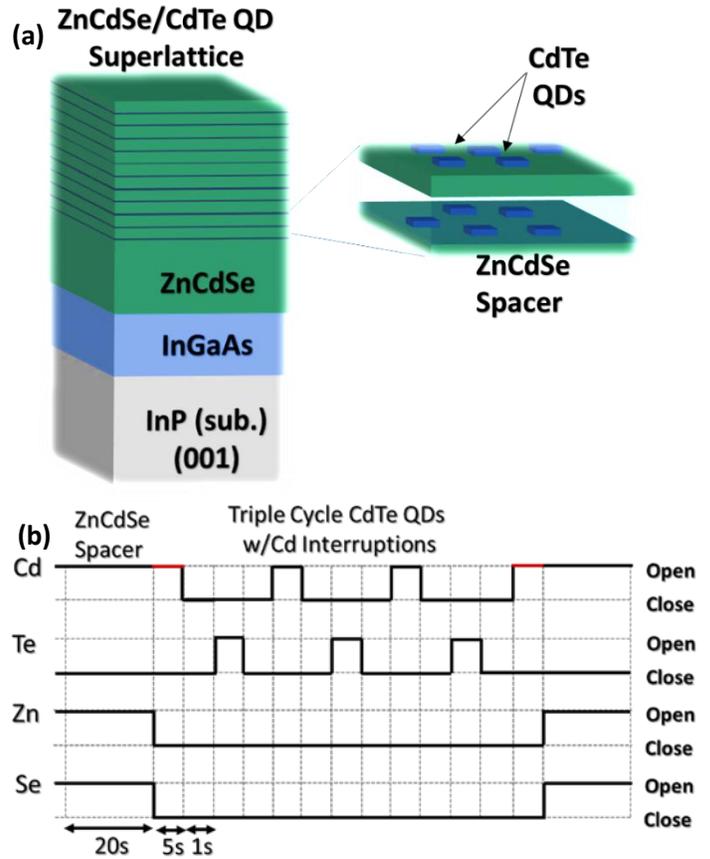

**Figure 1**. (**a**) Schematic of type-II sub-monolayr QD stacks, that would be implemented as an intermediate band region in an IBSC device. (**b**) Growth sequence used for the formation of QDs grown by migration enhanced epitaxy.

The shutter opening and closing cycles used are shown in Fig. 1(b). At the end of the MEE sequence the Cd shutters were opened for 5s before resuming the ZnCdSe spacer layer growth (by opening the Se and Zn shutters). It should be noted that in this sequence, in contrast to the previous sequence (e.g., Refs. 12) the Zn shutter remains closed throughout the entire MEE cycle, further minimizing the likelihood of any unintentional ZnSe being formed during the process.

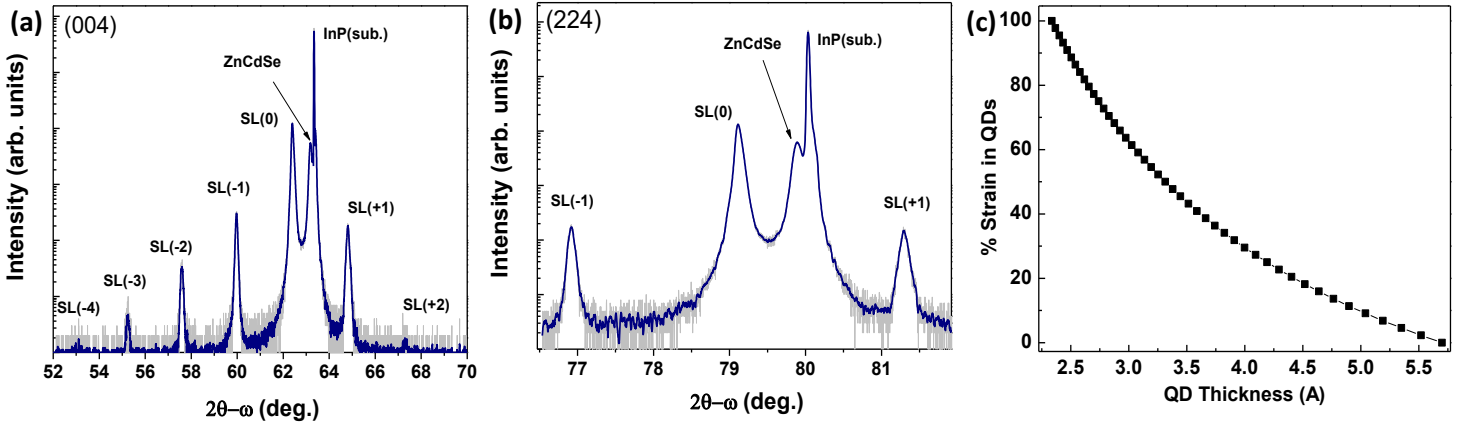

**Figure 2 (a)** HR-XRD along the (004) reflection; **(b)** The (224) asymmetric reflection. In conjunction, these two plots allow us to accurately measure the size and strain in the QDs. **(c)** Thickness of the QDs necessary to position the SL(0) peak under high compressive strain with respect to the substrate as observed in HR-XRD.

**Results:**

In Figs. 2(a) and 2(b) a high-resolution X-Ray diffraction (HR_XRD), for symmetric (004) and asymmetric (224) Bragg reflections, respectively, 2θ-ω scans are presented. The Bragg diffraction peaks of the different layers are clearly resolved. They include (see Fig. 2(a)): the ZnCdSe buffer layer (63.18°), the zeroth (0$^{th}$)-order superlattice peak SL(0) (62.40°) as well as higher order satellite peaks, which originate from the periodic CdTe/ZnCdSe QD SL structure. Such strong and sharp higher order satellite peaks suggest high quality materials and interfaces: well-controlled separation of QD layers and well-contained spacer segregation of materials. By combining the symmetric (004) and asymmetric (224) Bragg reflections the strain and composition of the different layers within the structure can be accurately calculated.

The ZnCdSe buffer layer has an in-plane lattice parameter, $a_\parallel^{Sp}$, that is equal to that of the InP substrate, $a_\parallel^{Sub} = 5.869$ Å. The layer is under slight compressive strain, as its out-of-plane lattice parameter, $a_\perp^{Sp} = 5.882$ Å, is slightly larger. The in-plane and out-of-plane lattice parameters for the SL(0) are, respectively, $a_\parallel^{SL} = 5.882$ Å and $a_\perp^{SL} = 5.948$ Å. These values suggest that the superlattice region is nearly pseudomorphic to the InP substrate, due to the small difference between the in-plane lattice parameters which is only 0.22%. The SL period is made up of the combination of a spacer layer of the same composition as that of the ZnCdSe buffer layer and the CdTe QD layer, which we will show can be modelled as fractional layers of CdTe. For a more complete understanding of the energy at which the IB level forms within the host material, a good understanding of the strain and the thickness of the quantum dots is necessary. For this we utilize the symmetric and anti-symmetric HR-XRD scans (Figs. 2(a) and 2(b)) using the following simple arguments and considerations.

From the (004) scan the out-of-plane thickness of the period can be obtained by the distance between higher order superlattice peaks and is calculated to be $t_\perp^{SL} = 43$Å. Considering the superlattice structure as a pseudo-crystal, we describe the lattice constant $a_\perp^{SL}$ as the weighted average of the strained individual layers that make up the period:

$$a_\perp^{SL} = \frac{a_\perp^{QDs} t_\perp^{QDs} + a_\perp^{Sp} t_\perp^{Sp}}{t_\perp^{SL}} \quad (1)$$

where, $t_\perp^{Sp}$, and $t_\perp^{QD}$ are the thickness of the ZnCdSe spacer and the average thickness of effective CdTe QD layer, respectively; $a_\perp^{QD}$ and $a_\perp^{Sp}$ are the strained lattice constants of the individual materials.

Fig. 2(c) relates the QD thickness and amount of strain imparted on the QDs necessary to match the measured out-of-plane lattice parameter of the SL from Eq. (1). For the unstrained case (0% strain) a significantly thicker dot is necessary, more than double than for the fully strained case (100% strain). By using Eq. (2), which relates the out of-plane-lattice parameter to the elastic constants of the SL, we can accurately calculate the thickness and strain of the QDs for this sample[17]:

$$a_\perp^{SL} = \left(1 + \frac{2C_{12}^{SL}}{C_{11}^{SL}}\right)(a_0^{SL} - a_\parallel^{SL}) + a_\parallel^{SL} \quad (2)$$

Here $C_{12}^{SL}$ and $C_{11}^{SL}$ are the weighted averages of the elastic constants of the individual unstrained CdTe and ZnCdSe layers given by[18]:

$$C_{ij}^{SL} = \frac{C_{ij}^{QDs} t_0^{QDs} + C_{ij}^{Sp} t_0^{Sp}}{t_0^{SL}} \quad (3)$$

where $t_0^{QDs}$, $t_0^{Sp}$, $t_0^{SL}$ are thickness of the unstrained corresponding layers. These thicknesses, as well as the thickness of strained layers can be conveniently expressed in terms of the number of corresponding monolayers, $N^{(i)}$, which stay constant between the strained and unstrained cases:

$$t_{\perp,0}^{SL} = \frac{a_{\perp,0}^{QDs}}{2} N^{QDs} + \frac{a_{\perp,0}^{Sp}}{2} N^{Sp} \quad (4)$$

Correspondingly,

$$a_{0,\perp}^{SL} = \frac{a_0^{QDs} \frac{a_\perp^{QDs}}{2} N^{QDs} + a_0^{Sp} \frac{a_{0,\perp}^{Sp}}{2} N^{Sp}}{t_{0,\perp}^{SL}} \quad (5)$$

We note here that using Eq. (4), weighted averages for elastic constants are also expressed in terms of number of monolayers and lattice constants, instead of layer thicknesses.

In Eq. 5 important parameters that are to be considered are $a_0^{Sp}$ and $a_\perp^{Sp}$. The former is known, since the composition of the spacer is the same as that of the buffer; however, the latter is not since we don't know the degree of relaxation, if any, in the spacers. We, nevertheless, know the range in which the in-plane lattice parameter of the spacer, $a_\parallel^{Sp}$ can vary. This is between $a^{Sub}$ (fully strained) and $a_0^{Sp}$ (fully relaxed). Therefore, we will take this, in further calculations, as a varying parameter, which, as shown below, uniquely determines the unknowns $(a_\perp^{QDs}, N^{QDs}, t_\perp^{QDs})$ for our given case via the measured SL lattice parameters and period.

Using Eqs. (4) and (5) we can get expressions for number of monolayers as follows:

$$N^{QD} = 2t_\perp^{SL} \frac{(a_\perp^{SL} - a_\perp^{Sp})}{a_\perp^{QD}(a_\perp^{QD} - a_\perp^{Sp})} \quad (6)$$

$$N^{Sp} = 2t_\perp^{SL} \frac{(a_\perp^{QD} - a_\perp^{SL})}{a_\perp^{Sp}(a_\perp^{QD} - a_\perp^{Sp})} \quad (7)$$

Inserting Eqs. (6), (7) and (3) back into Eq. (2) an expression in which $a_\perp^{QD}$ is the only unknown parameter, can be formulated:

$$(a_\perp^{SL} - a_\parallel^{SL}) = \left(1 + 2\frac{\beta a_\perp^{Sp} C_{12}^{QDs} + x a_\perp^{Sp} C_{12}^{Sp}}{\beta a_\perp^{Sp} C_{11}^{QDs} + x a_0^{Sp} C_{11}^{Sp}}\right) \times \quad (8)$$

$$\left(\frac{\alpha a_0^{QDs} + x a_0^{Sp}}{\alpha + x} - a_\parallel^{SL}\right)$$

Here $x = a_\perp^{QD}(a_\perp^{QD} - a_\perp^{SL})$, $\alpha = a_0^{QDs}(a_\perp^{SL} - a_0^{Sp})$, $\beta = a_0^{QDs}(a_\perp^{SL} - a_\perp^{Sp})$; $a_\perp^{QDs}$ can be calculated from $a_\parallel^{Sp}$ via an equation similar to Eq. (2). Eq. (8) can be solved for $a_\perp^{QDs}$ and ultimately for $N^{QDs}$, $N^{Sp}$ and the thicknesses of both the QDs and the spacer.

The out-of plane lattice parameter for the QDs is calculated to be $a_\perp^{QDs} = 7.037$ Å. This value does not change much when varying the spacer in-plane lattice parameter (effectively changing the strain) within the range we are working in. But, the fractional coverage of the CdTe QDs changes in our calculation significantly, as illustrated in Figure 3. The effective number of monolayers varies from 0.66 to 0.76. In both scenarios (strained and unstrained spacer) the average thickness is submonolayer in quantity for the CdTe QDs.

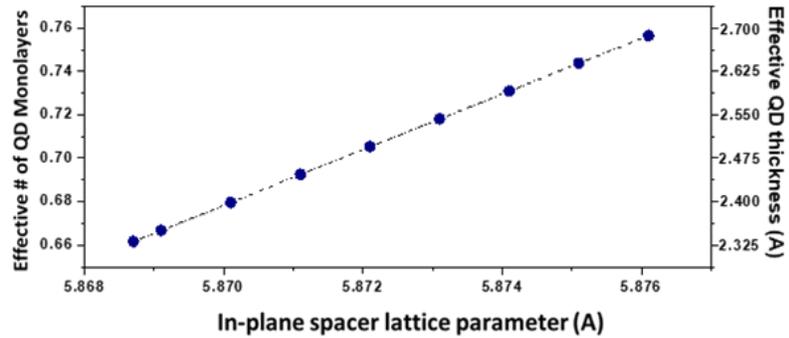

**Figure 3** QD height (in monolayers) and effective QD thickness are plotted with respect to the strain of the spacer. Sub-monolayer quantities are found to explain the experimental data.

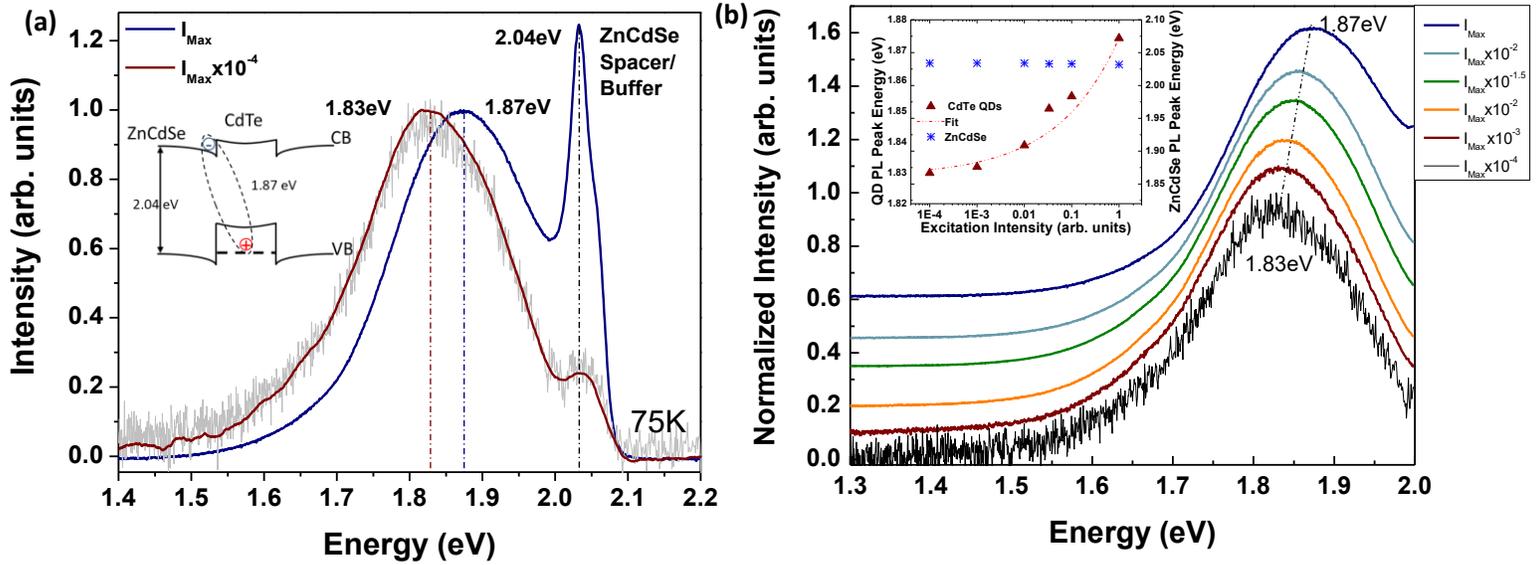

**Figure 4 (a)** Low temperature intensity dependent PL from the QDs as well from the ZnCdSe spacer and buffer region. The inset illustrates the type-II band alignment in which holes are confined within the QDs and electrons within ZnCdSe spacer. **(b)** Intensity dependent PL measurement confirms the type-II band alignment between QDs and host material. (inset) Plot of peak position of the QD PL as a function of excitation intensity, exhibiting a 1/3 power law fit. The energy of the ZnCdSe peak is plotted for comparison.

In order to investigate the suitability of the structure for an IBSC, we measured its luminescent properties. Figure 4a shows the PL spectrum of the structure taken at 75 K. PL emission was acquired by exciting the sample using a 50 mW 405 nm diode laser and the collected PL was analyzed with a HR4000 Ocean Optics spectrometer. In figure 3a two peaks can be identified from the spectrum. A sharp peak around 2.04 eV is due to the ZnCdSe buffer and spacer layers. The difference in energy from the bandgap energy of a lattice matched $Zn_{0.51}Cd_{0.49}Se$ to InP (around 2.1 eV) is to be expected since the buffer layer was grown with a slightly Cd rich composition, as indicated by the XRD[19]. A broad peak observed at lower energies, centered around 1.86 eV, is due to the CdTe QDs. Such a broad peak is reasonable due to expected size distribution and the type-II nature of the recombination process. Juxtaposed is a spectrum taken at a lower excitation intensity and a noticeable shift to lower energy of the QD peak by as much as 40 meV is observed.

Intensity dependent PL shown in Figure 4(b) was taken to verify the type-II band alignment for the low energy peak. In a type- II heterostructure, with higher excitation intensity an increased flux of electrons is promoted into the conduction band, producing a band bending effect at the interface between the QDs and the host material. The staggered band alignment between the two materials forms a triangular potential well at this interface in which the energy scales as a function of the cube root of the excitation intensity[20]. In the inset in figure 4(b), the peak position shift is plotted for the QDs and the ZnCdSe spacer. There is a clear shift in energy with excitation power for the QD PL that fits very well with a 1/3 power law fit, whereas the peak of the maximum excitation position of the ZnCdSe does not change with intensity. At the lowest excitation intensity, assuming a flat band potential between the QDs and the host material, a difference in energy of ~200 meV is observed between the band to band transition in the barrier layer (2.04 eV) and the type II transition arising from the QDs. This value suggests that the QDs in our structure are very small, thus shifting the confined hole energy level to near the VB edge of the ZnCdSe spacer.

Using Vegard's law and the known band alignments for these two materials[21,22], the band structure for the QD region is approximated in figure 5(a). There is a large valence band offset of 0.86 eV between ZnCdSe and CdTe and the large quantum confinement of the holes due to their very small size gives rise to the observed 200 meV energy difference. Increasing the size of the dots will allow the confined energy level of the QDs to shift closer to the CdTe valence band edge, thus reducing the type-II transition energy and increasing the energy difference between the QDs and the host material to the desired 0.5 – 0.7 eV range based on the device design. The large VBO of 0.86 eV provides sufficient tunability of the confinement energy in order to achieve these values. However, since the heavy hole energy level also depends strongly on the strain of the QD within the host material, strain effects must be

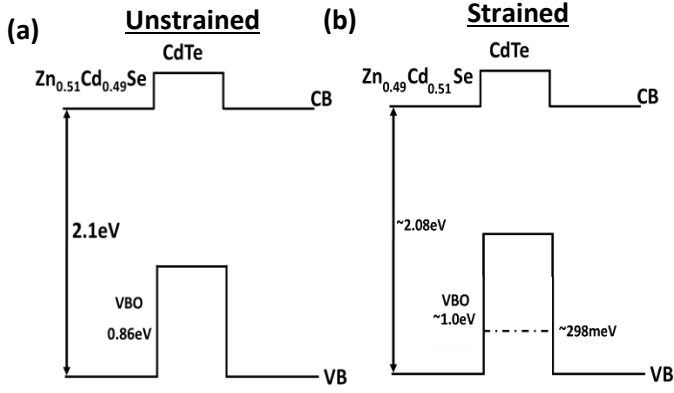

**Figure 5 (a)** Schematic band diagram between unstrained CdTe and ZnCdSe lattice matched to InP. **(b)** Band energy diagram from calculated values of composition ascertained from HR-XRD and taking into account for effects of strain.

considered. Due to the high lattice mismatch, between ZnCdSe and CdTe, biaxial compressive strain gives rise to hydrostatic strain component and to a first order approximation there is a splitting of the light and heavy hole levels due to the shear component of the strain that can be calculated by standard deformation potential theory.[23] The equation below calculates the new heavy hole (HH) energy due to strain,

$$\Delta E_{hh} = a_v(\varepsilon_{xx} + \varepsilon_{yy} + \varepsilon_{zz}) + \frac{b}{2}(\varepsilon_{xx} + \varepsilon_{yy} - 2\varepsilon_{zz}) \quad (9)$$

Where $a_v$ is the linear hydrostatic deformation potential for the valence band maxima and $b$ is the shear uniaxial deformation potential. The strain components are given by,

$$\varepsilon = \varepsilon_{xx} = \varepsilon_{yy} = \frac{a_{Sub} - a_{QDs}}{a_{Sub}} \quad (10)$$

and

$$\varepsilon_{zz} = \frac{-2C_{12}}{C_{11}}\varepsilon \quad (11)$$

Typically, compressive strain increases the band gap of the material and, depending on how large the strain is, it could significantly increase the VBO in a heterostructure. In our case this is highly desirable and allows for another tuning parameter to obtain the idealized value for the IB band. Due to the large lattice mismatch between the CdTe QDs and the spacer regions a large shift in the heavy hole energy can ultimately be observed that would be beneficial in being able to tune the energy level of the IB band. From our calculation in HR-XRD we find that the QDs are partially relaxed (35%). Thus, we find the VBO is now ~ 1.0 eV. The HH energy is calculated, using self-consistent variational calculation of the Schrödinger equation,[24] for a single monolayer of CdTe under these strain conditions, to be 289 meV, illustrated in figure 5(b). This is in close agreement of our PL results. Below is a list of the materials parameters used for these calculations.

**Table I.**      **Material Parameters used in calculations**

|       | $c_{11}$ (x10$^{10}$ Pa) | $c_{12}$ (x10$^{10}$ Pa) | $\alpha_v$ (eV) | $\beta_v$ (eV) |
|-------|--------------------------|--------------------------|-----------------|----------------|
| ZnSe  | 8.26 [25]                | 4.98 [25]                | 1.23 [26]       | -1.20 [26]     |
| CdSe  | 6.67 [27]                | 4.63 [27]                | 0.90 [25]       | -1.26 [27]     |
| CdTe  | 5.62 [25]                | 3.94 [25]                | 0.89 [26]       | -1.20 [25]     |

**Conclusion:**
We have successfully grown submonlayer CdTe QDs within a ZnCdSe host. The structural quality of the material was confirmed by the HR-XRD and their optical properties were established using intensity dependent PL. These materials have interesting properties that could be used towards the design of an ideal IBSC. The large VBO offset that exists between CdTe and the ZnCdSe host material, which is sufficient to provide the needed IB energy values, can be further tuned with strain. To have an accurate understating of the strain of the QDs we used HR-XRD to analyze the superlattice structure. We found that the dots were partially relaxed, and their fractional coverage could be extracted from the strain measured within the spacer region. Even though the dots were partially relaxed, our calculations show that we should expect an increase in the VBO of about 140 meV due to strain, which is helpful towards tuning the IB energy level to the desired value of 0.7eV. From the PL measurements and energy calculations we concluded that the dots in the current structure are too small (~1 monolayer in height) and thus the growth must be modified to achieve larger (thicker) dots.

**Acknowledgements:** This work was supported by NSF CBET award number 1512017 and NSF award number HRD-1547830 (CREST-IDEALS).

**References:**

[1] W. Shockley, H.J. Queisser, Detailed balance limit of efficiency of p–n junction solar cells, Journal of Applied Physics 32 (1961) 510519.

[2] A. Luque, A. Martí, Increasing the efficiency of ideal solar cells by photon induced transitions at intermediate levels, Physical Review Letters 78 (1997) 5014–5017.

[3] A. Martí, L. Cuadra, A. Luque, Quantum dot intermediate band solar cell, Conference Record of the 28th IEEE Photovoltaics Specialists Conference, New York, USA, (2000) 940-943.

[4] A. Luque, A. Martí, E. Antolín, C. Tablero, Intermediate bands versus levels in non-radiative recombination, Physica B 382 (2006) 320–327.

[5] I. Ramiro, J. Villa, C. Tablero, E. Antolín, A. Luque, A. Martí, Analysis of the intermediate-band absorption properties of type-II GaSb/GaAs quantum-dot photovoltaics, Physical Review B 96 (2017) 125422.

[6] (6) S.M. Hubbard, C.D. Cress, C.G. Bailey, R.P. Raffaelle, S.G. Bailey, D.M. Wilt, Effect of strain compensation on quantum dot enhanced GaAs solar cells, Applied Physics Letters 92, (2008) 123512.

[7] L. Zhu, K.H. Lee, M. Yamaguchi, H. Akiyama, Y. Kanemitsu, K. Araki, N. Kojima, Analysis of nonradiative recombination in quantum dot solar cells and materials, Progress in Photovoltaics: Research and Applications (2019), doi.org/10.1002/pip.3110.

[8] S. Dhomkar, U. Manna, L. Peng, R. Moug, I.C. Noyan, M.C. Tamargo, I.L. Kuskovsky, Feasibility of submonolayer ZnTe/ZnCdSe quantum dots as intermediate band solar cell material system, Solar Energy Materials & Solar Cells, 117 (2013) 604-609.

[9] S. Dhomkar, U. Manna, I.C. Noyan, M.C. Tamargo, I.L. Kuskovsky, Vertical correlation and miniband formation in submonolayer Zn (Cd) Te/ZnCdSe type-II quantum dots for intermediate band solar cell application, Applied Physics Letters 103 (2014)18, 181905.

[10] C.M. Imperato, G.A. Ranepura, L.I. Deych, I.L. Kuskovsky, Theoretical Determination of Optimal Material Parameters for ZnCdTe/ZnCdSe Quantum Dot Intermediate Band Solar Cells, Journal of Electronic Materials, 47 (2018) 4325–4331.

[11] M.Y. Levy, C. Honsberg, A. Marti, A. Luque, Quantum Dot Intermediate Band Solar Cell Material Systems with Negligible Valence Band Offsets, Conference Record of the IEEE Photovoltaic Specialists Conference (2005) 90-93.

[12] S. Dhomkar, I.L. Kuskovsky, U. Manna, I.C. Noyan, M.C. Tamargo, Optimization of growth conditions of type-II Zn (Cd) Te/ZnCdSe submonolayer quantum dot superlattices for intermediate band solar cells, Journal of Vacuum Science & Technology B 31(3) (2013).

[13] A. Marti, N. López, E. Antolín, E. Cánovas, A. Luque, C. R. Stanley, C. D. Farmer, and P. Díaz, Emitter degradation in quantum dot intermediate band solar cells, Applied Physics Letters 90 (2007) 233510.

[14] V. Popescu, G. Bester, M.C. Hanna, A.G. Norman, A. Zunger, Theoretical and Experimental Examination of the Intermediate-Band Concept for Strain-Balanced (In,Ga)As/Ga(As,P) Quantum Dot Solar Cells, Physical Review B 78 (2008) 20532.

[15] A. Takata, R. Oshima, Y. Shoji, K. Akahane, Y. Okada, Fabrication of 100 layer-stacked InAs/GaNAs strain-compensated quantum dots on GaAs (001) for application to intermediate band solar cell, 35th IEEE Photovoltaic Specialists Conference (2010).

[16] V. Deligiannakis, S. Dhomkar, M.S. Claro, I.L. Kuskovsky, M.C.Tamargo Interface Modification in Type-II ZnCdSe/Zn(Cd)Te QDs for High Efficiency Intermediate Band Solar Cells, Journal of Crystal Growth 512 15 (2019) 203-207.

[17] T. Rieger, T. Riedl, E. Neumann, D. Grützmacher, J. Lindner, A. Pawlis, Strain Compensation in Single ZnSe/CdSe Quantum Wells: Analytical Model and Experimental Evidence, ACS Applied Materials & Interfaces 9(9) (2017).

[18] S.J. Polly, C.G. Bailey, A.J. Grede, D.V. Forbes, S.M. Hubbard, Calculation of strain compensation thickness for III–V semiconductor quantum dot superlattices, Journal of Crystal Growth 454 (2016).

[19] A. Sweiti, F. Medina, L. Martinez, A. Lopez-Rivera, Photoluminescence spectroscopy and effective concentration determination of $Cd_xZn_{1-x}Se$, Semiconductor Science and Technology, 23(3) (2008) 035019.

[20] F. Hatami, M. Grundmann, N. N. Ledentsov, F. Heinrichsdorff, R. Heitz, J. Böhrer, Z.I. Alferov, Carrier dynamics in type-II GaSb/GaAs quantum dots, Physical Review B, 57(8) (1998) 4635–4641.

[21] M.C. Tamargo, II-VI Semiconductor Materials and their Applications (Optoelectronic Properties of Semiconductors and Superlattices) 1st Edition, CRC Press (1997)

[22] M. Ściesiek, J. Suffczyński, W. Pacuski, M. Parlińska-Wojtan, T. Smoleński, P. Kossacki, A. Golnik, Effect of electron-hole separation on optical properties of individual Cd(Se,Te) quantum dots, Physical Review B 93, (2016) 195313.

[23] S.L. Chuang, Physics of Photonic Devices, 2nd ed. Hoboken: Wiley (2012) 132–142.

[24] I.V. Ponomarev, L.I. Deych, V.A. Shuvayev, A.A. Lisyansky, Self-consistent approach for calculations of exciton binding energy in quantum wells, Physica E 25 (2005) 539.

[25] Y.H. Li, X.G. Gong, S.H. Wei, Ab initio all-electron calculation of absolute volume deformation potentials of IV-IV, III-V, and II-VI semiconductors: The chemical trends, Physical Review B 73 (2006) 245206.

[26] C.G. Van de Walle, Band lineups and deformation potentials in the model-solid theory, Physical Review B 39 (1989) 1871.

(27) H. J. Lozykowski, V.K. Shastri, Excitonic and Raman properties of ZnSe/ZnCdSe strained layer quantum wells Journal of Applied Physics 69 (1991) 3235.